\documentclass[preprint,12pt]{elsarticle}

\usepackage{amssymb}
\usepackage{amsmath}
\usepackage{listings}
\usepackage{graphicx, caption, subcaption}
\usepackage{xcolor}
\usepackage{hyperref}

\colorlet{punct}{red!60!black}
\definecolor{background}{HTML}{EEEEEE}
\definecolor{delim}{RGB}{20,105,176}
\colorlet{numb}{magenta!60!black}

\lstdefinelanguage{json}{
    basicstyle=\normalfont\ttfamily,
    numbers=left,
    numberstyle=\scriptsize,
    stepnumber=1,
    numbersep=8pt,
    showstringspaces=false,
    breaklines=true,
    frame=lines,
    literate=
     *{0}{{{\color{numb}0}}}{1}
      {1}{{{\color{numb}1}}}{1}
      {2}{{{\color{numb}2}}}{1}
      {3}{{{\color{numb}3}}}{1}
      {4}{{{\color{numb}4}}}{1}
      {5}{{{\color{numb}5}}}{1}
      {6}{{{\color{numb}6}}}{1}
      {7}{{{\color{numb}7}}}{1}
      {8}{{{\color{numb}8}}}{1}
      {9}{{{\color{numb}9}}}{1}
      {:}{{{\color{punct}{:}}}}{1}
      {,}{{{\color{punct}{,}}}}{1}
      {\{}{{{\color{delim}{\{}}}}{1}
      {\}}{{{\color{delim}{\}}}}}{1}
      {[}{{{\color{delim}{[}}}}{1}
      {]}{{{\color{delim}{]}}}}{1},
}

\newcounter{bla}

\journal{Computer Physics Communications}

\begin{document}
\lstset{
    frame       = tb,
    numbers     = left,
    showspaces  = false,
    showstringspaces    = false,
    captionpos  = b,
    caption     = \lstname
}
\begin{frontmatter}



\title{PyQCAMS: Python Quasi-Classical Atom-Molecule Scattering}


\author[a,b]{Rian Koots}
\author[a,b]{Jesús Pérez-Ríos\corref{author}}

\cortext[author] {Corresponding author.\\\textit{E-mail address:} jesus.perezrios@stonybrook.edu}
\address[a]{Department of Physics, Stony Brook University, Stony Brook, New York 11794, USA}
\address[b]{Institute for Advanced Computational Science, Stony Brook University, Stony Brook, New York 11794, USA}

\begin{abstract}
We present Python Quasi-classical atom-molecule scattering (PyQCAMS), a new Python package for atom-molecule scattering within the quasi-classical trajectory approach. The input consists of mass, collision energy, impact parameter, and pair-wise interactions to choose between Buckingham, generalized Lennard-Jones, and Morse potentials. As the output, the code provides the vibrational quenching, dissociation, and reactive cross sections along with the rovibrational energy distribution of the reaction products. Furthermore, we treat H$_2$ + Ca $\rightarrow$ CaH + H reactions as a prototypical example to illustrate the properties and performance of the software. Finally, we study the parallelization performance of the code by looking into the time per trajectory as a function of the number of CPUs used. 

\end{abstract}


\end{frontmatter}

% \linenumbers

% Computer program descriptions should contain the following
% PROGRAM SUMMARY.

%{\bf PROGRAM SUMMARY/NEW VERSION PROGRAM SUMMARY}
  %Delete as appropriate.
% All CPiP articles must contain the following
% PROGRAM SUMMARY.

{\bf PROGRAM SUMMARY}
  %Delete as appropriate.

\begin{small}
\noindent
{\em Program Title: PyQCAMS: Python Quasi-Classical Atom-Molecule Scattering} \\
{\em CPC Library link to program files:} (to be added by Technical Editor) \\
{\em Developer's repository link:} \url{https://github.com/Rkoost/PyQCAMS} \\
{\em Code Ocean capsule:} (to be added by Technical Editor)\\
{\em Licensing provisions:} MIT\\
{\em Programming language: Python}                                   \\
% {\em Supplementary material:}                                 \\
  % Fill in if necessary, otherwise leave out.
%{\em Journal reference of previous version:}*                  \\
  %Only required for a New Version summary, otherwise leave out.
%{\em Does the new version supersede the previous version?:}*   \\
  %Only required for a New Version summary, otherwise leave out.
%{\em Reasons for the new version:*}\\
  %Only required for a New Version summary, otherwise leave out.
%{\em Summary of revisions:}*\\
  %Only required for a New Version summary, otherwise leave out.
{\em Nature of problem: Simulation of atom-molecule scattering systems.}\\
  %Describe the nature of the problem here. \\
{\em Solution method: Quasi-Classical trajectory method and numerical solution of Hamilton's equations.}\\
  %Describe the method solution here.
% {\em Additional comments including restrictions and unusual features (approx. 50-250 words):}\\
  %Provide any additional comments here.
\end{small}
   \\

\section{Introduction}
For decades, one of the main approaches to studying molecular dynamics has been via the quasi-classical trajectory (QCT) method \cite{karplus,truhlar1979}. This technique treats collisions semi-classically. The nuclear dynamics in the underlying potential energy surface is treated classically. However, the initial and final states are picked following the Bohr-Sommerfeld quantization rule, yielding accurate predictions at significantly less cost than quantum dynamics as long as they fall within given conditions regarding the collision energy --number of contributing partial waves \cite{bonnet1997,perezriosbook}. QCT has been used in multitude of scenarios relevant to chemical physics~\cite{Aoiz1992,Aoiz1998,Oliveira2014,Tibor2018,Petra2022,Meuwly2022,Goswami2022}, and cold and ultracold chemistry~\cite{perezrios_rydberg,PerezRios2019,Hirzler2020}, ranging from the ultracold to the hyper-thermal regimes. In particular, It has been used to study the relaxation and reaction dynamics of cold atom-ionic molecule \cite{PerezRios2019,perezrios_rydberg} and atom-molecule collisions \cite{mota2022}. 

We present an open-source, object-oriented program written in Python to perform QCT calculations on atom-molecule systems called Python Quasi-Classical Atom-Molecule Scattering (PyQCAMS). While chemical dynamics programs such as Gaussian \cite{g16} and VENUS \cite{hase1996general}, we introduce a more accessible, user-friendly and dedicated path to performing these simulations. Our program consists of completely open-source software, and relies mainly on the NumPy \cite{numpy} and SciPy \cite{scipy} packages. We use matplotlib\cite{matplotlib} for data visualization,  pandas\cite{pandas} for data storage and analysis, and multiprocess \cite{multiprocess1,pathos} for parallel implementation. 

The outline of this paper is as follows: In Section \ref{theory} we discuss the theory behind the quasi-classical trajectories method, including initial conditions, trajectory reactions, and analysis.  In Section \ref{program}, we discuss the PyQCAMS program as separated into the inputs, main code, and outputs. We also discuss the implementation and performance of the program. In Section \ref{example}, we provide an example of a typical workflow to study the reaction (H$_2$ + Ca), where we outline how a user can obtain reaction rates and product distributions using the program. 

\section{Theoretical approach}\label{theory}
\begin{figure}
    \centering
    \includegraphics{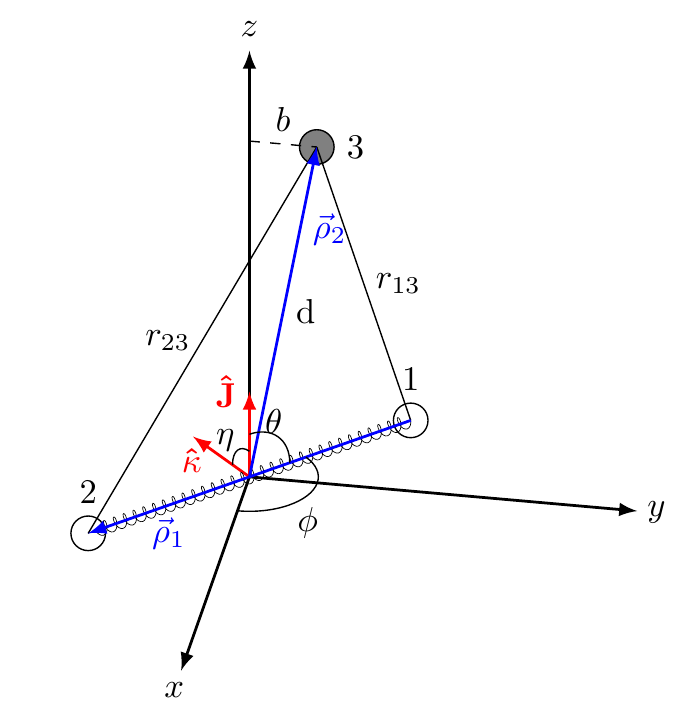}
    \caption{(Color online.) Jacobi coordinates of a atom-molecule system. Here, the molecule is rotating in the x-y plane, with the angular momentum $\vec{J}$ along the z-axis and $\vec{\kappa} = \vec\rho_1 \times \hat{z}$. $\eta$ is the angle between $\vec\kappa$ and $\vec{J}$, and $\phi$ and $\theta$ are defined as usual in spherical coordinates.}
    \label{fig:jacobi}
\end{figure}

In this section we describe the basics of QCT. 
The quasi-classical trajectory (QCT) method treats scattering processes semi-classically. First, by solving Newton's equations of motion of the colliding nuclei. Next, in the case of atom-molecule scattering, at the start of each trajectory, the internal degrees of freedom of the molecule are treated within the Wentzel-Kramers-Brillouin (WKB) approximation, such that the initial rovibrational state ($v,j$) satisfies a quantum-mechanically viable state. The classical Hamiltonian for a 3-particle, atom-molecule system with masses $m_i$, $i = 1,2,3$ takes the form:
\begin{equation}
    H = \sum_{i=1}^{3}\frac{\vec{p}_i^2}{2m_i} + V(\vec{r}_1,\vec{r}_2,\vec{r}_3)
\end{equation}
where $\vec{p}_i$ and $\vec{r}_i$ represent the momentum and position vectors of each atom with respect to the origin.
This Hamiltonian is expressed in Jacobi coordinates as:
\begin{equation}
    H = \frac{\vec{P}_1^2}{2\mu_{12}} + \frac{\vec{P}_2^2}{2\mu_{123}} + V(\vec\rho_1,\vec\rho_2) 
\end{equation}
where $\vec\rho_1$ is the Jacobi vector associated with the molecule and $\vec{P}_1$ is its conjugate momentum. $\vec\rho_2$ is the Jacobi vector connecting the atom with the center of mass of the molecule, and $\vec{P}_2$ is its conjugate momentum, as shown in Fig.~\ref{fig:jacobi}. Note that this Hamiltonian uses the reduced masses of the corresponding atoms: $\mu_{12} = (\frac{1}{m_1} + \frac{1}{m_2})^{-1}$ and $\mu_{123} = (\frac{1}{m_3} + \frac{1}{m_{1}+m_{2}})^{-1}$. Defining the coordinates in this way separates the center-of-mass degree of freedom from the relative one. Then, the momentum associated with the  center-of-mass degrees of freedom is a constant of motion since the interaction potential do not depend on the center-of-mass position, and neglected in the analysis of the dynamics.. 

We trace the atoms' subsequent motion by solving Hamilton's equations of motion:
\begin{equation}
    \frac{d\rho_{i,\alpha}}{dt} = \frac{\partial H}{\partial P_{i,\alpha}}
\end{equation}
\begin{equation}
    \frac{dP_{i,\alpha}}{dt} = -\frac{\partial H}{\partial\rho_{i,\alpha}}
\end{equation}
for $i = 1,2$ for each associated vector and $\alpha = 1,2,3$ for each Cartesian component.
\subsection{Initial Conditions}\label{sec: icond}

The initial orientation angles are randomly generated to initialize each trajectory. The momentum of the molecule, $\vec{P}_1$, is subsequently defined by the orientation angles. If we initialize the molecule at its classical outer turning point $|\vec\rho_1| = r_+$, the momentum has no radial component. Since the angular momentum $\vec{J} = \vec{r}\times\vec{p}$, and the fact that $\vec\rho_1 \perp \vec{P}_1$, we find the magnitude of $P_1 = \hbar \sqrt{j(j+1)}/r_+$, with components \cite{perezriosbook}:
\begin{equation}
    \vec{P}_1 = 
    P_1
    \begin{pmatrix}
    \sin\phi\cos\eta - \cos\theta\cos\phi\sin\eta \\
    -\cos\phi\cos\eta - \cos\theta\sin\phi\sin\eta \\
    \sin\theta\sin\eta
    \end{pmatrix}
\end{equation}      

The initial vibrational phase is randomly generated by choosing the initial distance between the atom and molecule as
\begin{equation} \label{eq: dist}
    R = R_0 + \frac{\chi P_2\tau_{v,j}}{\mu_{123}}
\end{equation}
where $R_0$ is a fixed far-away distance where the interaction potential strength is negligible. The second term probes the molecule's vibrational phase since $\chi \in [0,1]$ is randomly generated following a uniform distribution and $\tau_{v,j}$ is the vibrational period of the molecule corresponding to the quantum-mechanical state ($v,j$). This is calculated as
\begin{equation}
    \tau_{v,j} = \sqrt{2\mu_{12}}\int_{r_-}^{r_+}\left[E_{int} - V(r_{12}) - \frac{\hbar^2j(j+1)}{2\mu_{12}r_{12}^2}\right]^{-\frac{1}{2}}dr
\end{equation}
where $V(r_{12})$ is the molecular potential energy and $r_{12}$ is the molecular separation.  $E_{int}$ is the internal energy of the molecule, which is calculated following a discrete variable representation (DVR) method using particle-in-a-box eigenfunctions \cite{colbert_miller_dvr}. 

\subsection{Reaction Products}
For the atom-molecule reaction AB + C, we expect three possible final products:
\begin{enumerate}
    \item Inelastic collision (quenching): AB($v$) + C $\rightarrow$ AB($v'$) + C 
    \item Molecular formation (reaction): AB + C $\rightarrow$ AC + B or BC + A 
    \item Dissociation: AB + C $\rightarrow$ A + B + C
\end{enumerate}
\begin{figure}
    \centering
    \includegraphics[width = \textwidth]{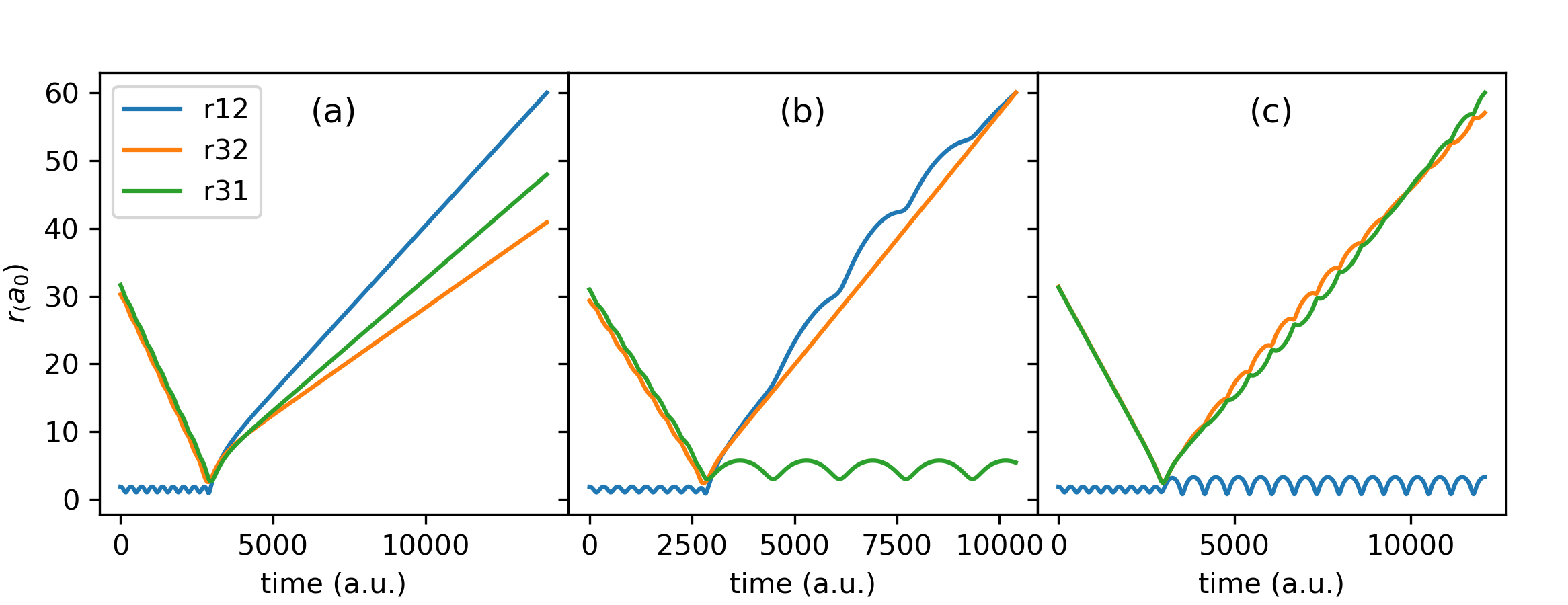}
    \caption{(Color online.) Different product outcomes for a given QCT calculation for H$_2$ + CaH at $E_c = 50000$ K, b = 0 $a_0$. $r_{12}$ represents the internuclear distance between each H, whereas $r_{32}$ and $r_{31}$ represent the internuclear distance between the colliding atom Ca (3) and each molecular atom (H(1) and H(2)). These trajectories show dissociation (a), reaction (b), and quenching (c). In panel (b), $r_{31}$ oscillates around a fixed distance, an indicator that a new molecule is formed.}
    \label{fig:trajectories}
\end{figure}

The final product is determined by the relative energy between each atom. The condition for whether two atoms are bound is determined by the effective potential:
\begin{equation}
    E_{ab} < 
    \begin{cases}
    V_{ab}(r_0)+\frac{j'(j'+1)}{2\mu_{ab} r_0^2} & j' \neq 0 \\
    0 & j' = 0
    \end{cases}
\end{equation}
where $r_0$ is the local maximum of the effective potential. Here, $a,b \in [1,3]$ represent each of the atoms. Two atoms are considered bound only if this condition is satisfied for just one pair of atoms, so that the other two pairs can be deemed unbound. 

In the same manner as the initial rovibrational states $(v,j)$, the final states $(v',j')$ are calculated within the WKB approximation. The rotational quantum number is given by:
\begin{equation}
    j' = -\frac{1}{2} + \frac{1}{2}\sqrt{1+4\frac{\vec{J'}\cdot{\vec{J'}}}{\hbar^2}}
\end{equation}
where $\vec{J} = \vec\rho_i \times \vec{P}_i$ is the effective conjugate angular momentum of the Jacobi vector $\vec\rho_i$. The vibrational quantum number is given by:
\begin{equation} \label{eq: vprime}
    v' = -\frac{1}{2} + \frac{\sqrt{2\mu_{ab}}}{\pi\hbar}\int_{r_-}^{r_+}\left[E'_{int} - V(r_{ab}) - \frac{\hbar^2j'(j'+1)}{2\mu_{ab}r_{ab}^2}\right]^{\frac{1}{2}}dr
\end{equation}
Although these equations lead to continuous numbers, they need to be interpreted as integers before assigning them as quantum numbers. This is done through the Gaussian binning (GB) process, where each $v'$ is weighted against its nearest integer $v_t$ with a Gaussian \cite{bonnet1997}:
\begin{equation}\label{eq: gauss}
    W_{q,r}(v',v_t) = \frac{1}{\sigma\sqrt{2\pi}}e^{-\frac{|v'-v_t|^2}{2\sigma^2}}
\end{equation}

\subsection{Analysis}
The cross section and reaction rates for an atom-molecule collision are calculated from QCT simulations. The first step in doing so is to calculate the opacity function, which is formally defined as
\begin{equation}
    P_{q,r,d}(E_c,b) = \int P_{q,r,d}(E_c,b,\theta,\phi,\eta,\chi){d}\Omega
\end{equation}
where $d\Omega = \sin\theta d\theta d\phi d\eta d\chi$. This integral is evaluated via the Monte Carlo technique over the randomly generated variables $\theta,\phi,\eta, \text{and } \chi$.
\begin{equation}
    P_{q,r,d}(E_c,b) = \frac{N_{q,r,d}(E_c,b)}{N} \pm \delta_{q,r,d}(E_c,b)
\end{equation}

The final state distribution of, for example, a reaction is calculated via the Gaussian Binning method \cite{bonnet2008}:
\begin{equation}
    P_r(E_c,b,v_t) = \frac{\sum_{i=1}^{N_r}W_r^i(v',v_t)}{W}
\end{equation}
with the total weight $W$ evaluated as
\begin{equation}
    W = \sum^{v_t}\left(\sum_i^{N_r}W_r^i(v',v_t) + \sum_i^{N_q}W^i_q(v',v_t)\right)  + N_d
\end{equation}
where $N_r,N_q, \text{and} N_d$ are the number of reaction, quenching, and dissociation products, respectively. Each Gaussian weight is summed over both the total number of products  and the total number of vibrational states produced by the scattering process. Each dissociation reaction has a weight $W_d = 1$ since there is no vibrational state associated with this product. 

For $N$ calculations, the opacity function $P_{q,r,d}(E_c,b)$ gives the probability of quenching (q), reacting (r), or dissociation (d) and $\delta$ is the error associated with the Monte Carlo technique:
\begin{equation}
    \delta_{q,r,d} = \frac{\sqrt{N_{q,r,d}(E_c,b)}}{N}\sqrt{\frac{N-N_{q,r,d}(E_c,b)}{N}}
\end{equation}

\section{The program}\label{program}
\begin{figure}
    \centering
    \includegraphics[width = \textwidth]{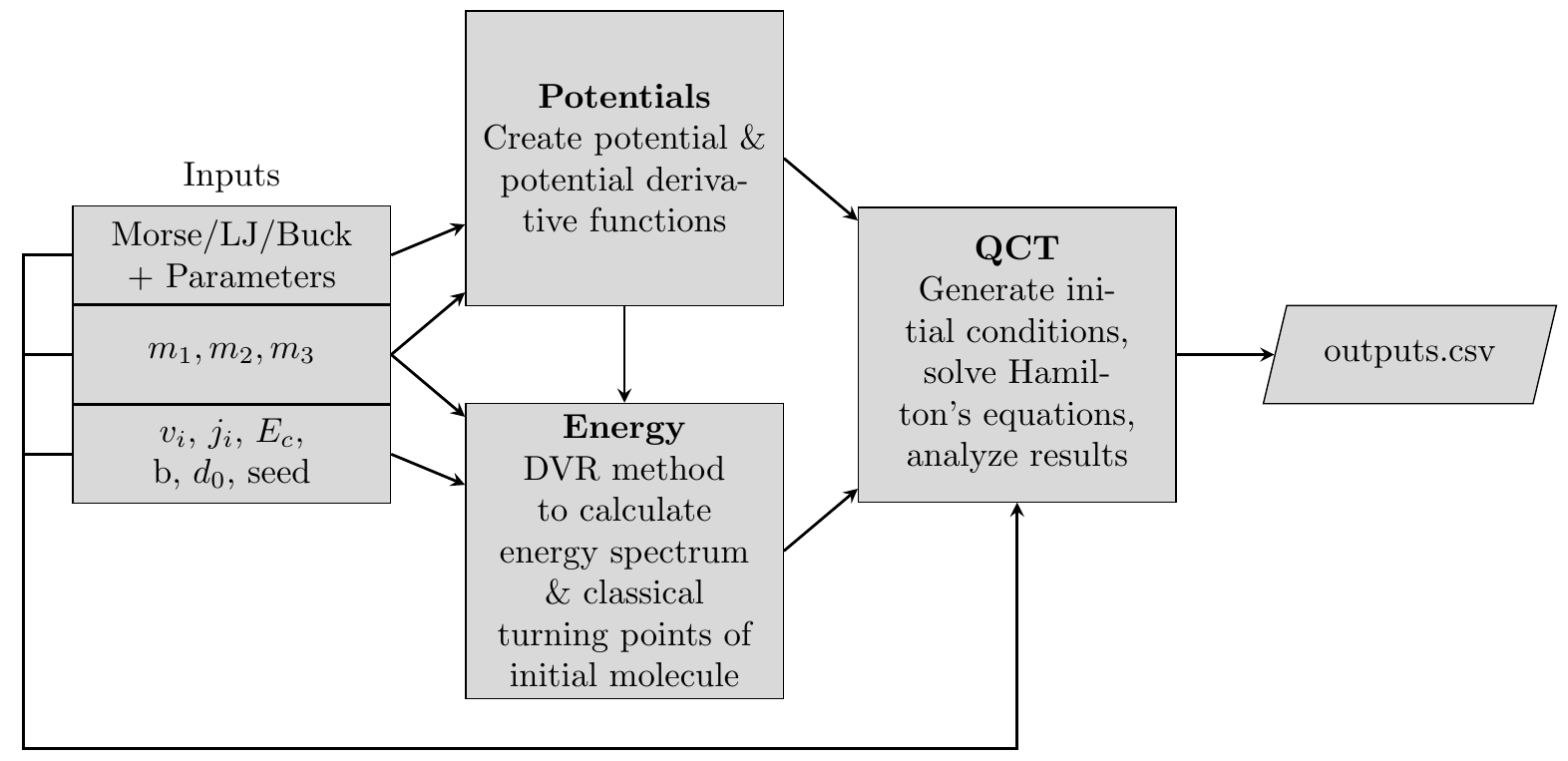}
    \caption{The general structure of the PyQCAMS program. }
    \label{fig:pyqcams_flow}
\end{figure}

PyQCAMS is written in an object-oriented manner, containing three main classes; \texttt{Potentials}, \texttt{Energy}, and \texttt{QCT}. This allows for the addition of new methods within each step of the calculation outlined in Figure \ref{fig:pyqcams_flow}. The program takes an input file where the user specifies the details of the reaction of interest. The variables are passed into their respective classes, where the trajectory calculations are performed. The results of each trajectory are then output into a \texttt{csv} file. The details of each step are outlined in this section.\\

\subsection{Input}
The input file is a JSON file whose keywords are described in this section (Listing \ref{code: input}). Each trajectory is tuned by the initial rovibrational energy level of the molecule \texttt{vi, ji}, the collision energy \texttt{Ec (K)}, impact parameter \texttt{b}, and initial distance between the atom and molecule \texttt{r0}. 
For reproducibility, the user can input a \texttt{seed} for the random number generator, or leave it as null to generate a new random number. The interaction potential between each atom is also tune-able, with choices from the Morse, generalized Lennard-Jones, and Buckingham potentials. The user should choose each of these internuclear potentials labeled as ``potential\_AB (BC,CA)". Note that AB should always represent the initial molecule, so that the ``C" represents the colliding atom. This convention should also be followed when specifying the masses.

The associated parameters of each of these potentials must be entered by the user, which are outlined in Section ~\ref{sec:main}. Since the energy spectrum is calculated via the DVR method, the user should also enter the number of DVR points and specify the range for which the potential is well-defined, such that it captures the potential minimum, short-range, and long-range behavior.

Finally, the parameters for integration should be entered. The user can choose when to stop a trajectory using \texttt{t\_stop}, which is a multiplicative factor of the trajectory's time scale. %sets $t_{\texttt{max}} = (d/v_c)t_{\texttt{stop}}$ 
Another stopping condition is the maximum distance between any two atoms, \texttt{r\_stop}. Finally, the absolute and relative tolerances of the integrator can be controlled with \texttt{a\_tol, r\_tol}.

Note that all values entered in the input file should be in atomic units except for collision energy, which is in K for the user's convenience. However, when importing the energy into the program, ensure that it is converted to atomic units, as is done in the \texttt{start()} method. 

\begin{lstlisting}[language = json,
                   basicstyle=\fontsize{9}{11}\ttfamily,
                   label = {code: input},
                   caption = {The input file for an H$_2$ + CaH trajectory calculation. Each parameter is explained in the text. The potential parameters for molecules BC and CA are collapsed to save space, but contain the same variables as those for AB.}]
    {"vi": 0, 
     "ji": 0,
     "Ec (K)": 30000,
     "b": 0,
     "r0": 30,
     "seed": null,
     "potential_AB": "morse",
     "potential_BC": "morse",
     "potential_CA": "morse",
     "masses":{"ma": 1837.47165336,
               "mb": 1837.47165336,
               "mc": 73046.7897752},
     "int_params":{"t_stop": 15,
                   "r_stop": 60,
                   "a_tol": 1E-10,
                   "r_tol": 1E-11},
     "potential_params":{
         "AB":{"npts": 1000,
               "xmin": 0.5,
               "xmax": 10,
               "morse":{"re" : 1.40104284795, 
                        "de" : 0.16456603489,
                        "we" : 0.02005340207},
               "lj":{"m":6,
                     "n":12,
                     "re" : 1.40104284795, 
                     "cm": 6.49902670540583931313,
                     "cn": 64.16474114146757},
               "buck":{"a": 167205.03207304262,
                       "b": 8.494089813101883, 
                       "c6": 6.49902670540583931313,
                       "re" : 1.40104284795,
                       "max": 0.2}
         },
         "BC": { ... },
         "CA": { ... }
         }
\end{lstlisting}
\subsection{Main Code}\label{sec:main}

The \texttt{Potentials} class houses the interaction potentials. Each method in this class represents a different potential, and returns a tuple of two function objects: the potential and its derivative. This class is useful for studying and manipulating the effect of different potentials on different pairs of the atom-molecule system. Results can be sensitive to the chosen potential, so it is important to test different potentials when studying a system. The available potentials and associated parameters are:
\begin{enumerate}
    \item Morse: $V(r) = D_e\left(1-\exp^{(-\alpha(r-r_e))}\right)^2 - D_e$ \\
    Parameters required: $D_e,r_e,w_e$
    \item Generalized Lennard-Jones: $V(r) = C_{m}/r^{m} - C_{n}/r^n$ \\
    Parameters required: $m,n, C_m, C_n$, and $r_e$ (guess left)
    \item Buckingham: $ V(r) = a e^{-br} - C_6/r^6 $ \\                 
    Parameters required:  $a,b, C_6$, and $r_e$ (guess left).
                ``max": Guess of where the maximum is. At short range, Buckingham potentials can reach a maximum and collapse. Enter your nearest $r$ value to this maximum.
                
\end{enumerate}
Note that while r$_e$ is not a parameter in the analytical expressions of the generalized Lennard-Jones and Buckingham potentials, it is required as part of a root solver method 
The \texttt{Energy} class houses the DVR method and turning point calculations. This is a separate class since these quantities should be computed once at the beginning of each trajectory. The spectrum and turning points for different E$_{(v,j)}$ in different potentials can be stored separately for future use.

\begin{figure}
    \centering
    \includegraphics[width = \textwidth]{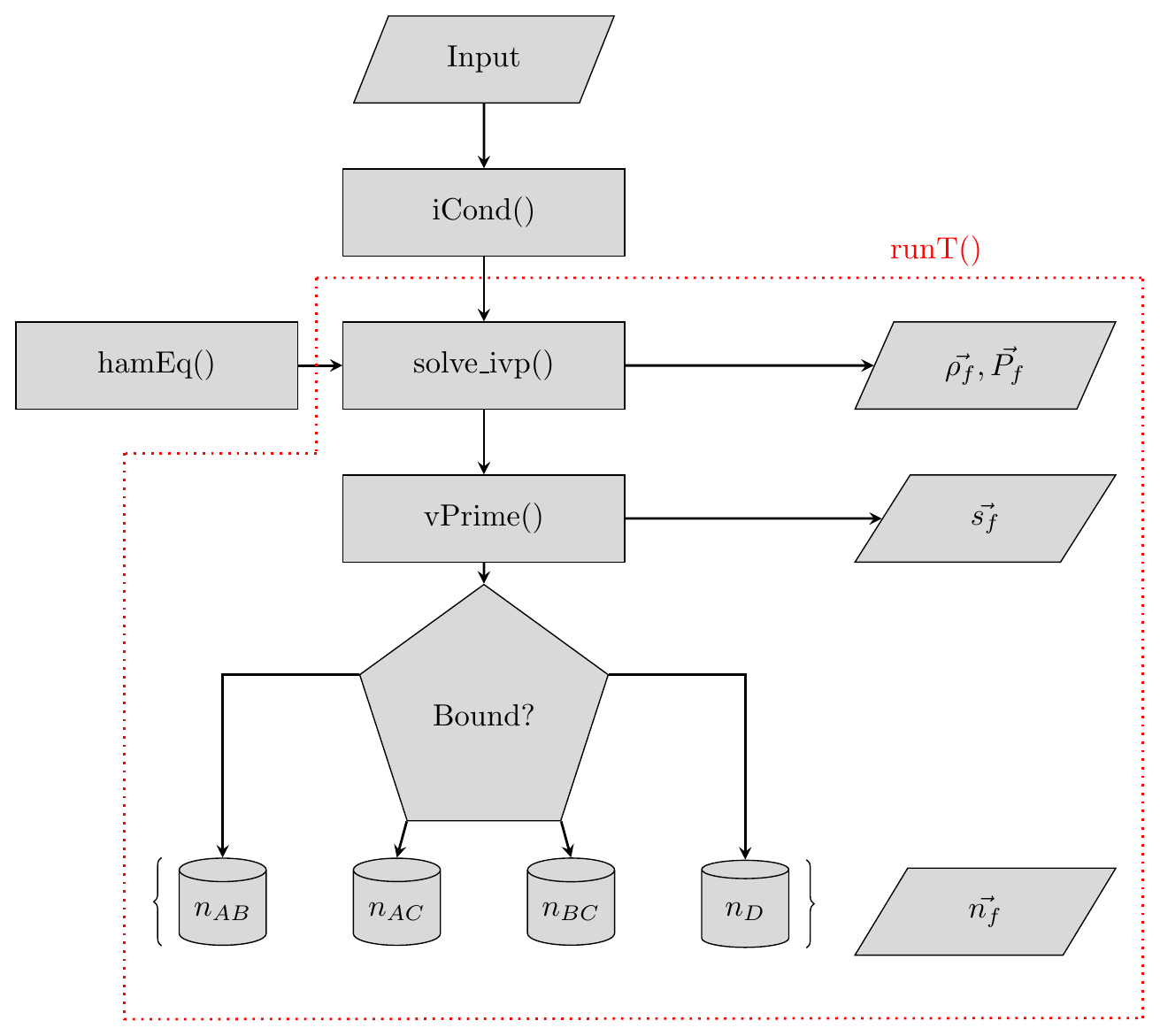}
    \caption{Outline of the \texttt{QCT} class. The components within the dashed line are contained in the \texttt{run\_T()} method of the \texttt{QCT} class. The outputs are the final position, momentum, state, and product count vectors. The product count vector is defined by the final product bins, filled according to which pair of atoms is bound.}
    \label{fig:qct_flow}
\end{figure}

The \texttt{QCT} class performs the main QCT calculation, and is outlined in Figure \ref{fig:qct_flow}. The \texttt{iCond()} method uses the calculated parameters from the \texttt{Energy} and \texttt{Potentials} class to generate a random set of initial conditions, yielding ($\vec{\rho_1}, \vec{\rho_2}, \vec{P_1}, \vec{P_1}$). The \texttt{hamEq()} function writes the Hamilton's equations of motion, and serves as an input function to \texttt{solve\_ivp()}, a \texttt{SciPy} \cite{scipy} integrator utilizing the adaptive Runge-Kutta 5(4) intregration method \cite{rungekutta}.  The \texttt{vPrime()} method serves to calculate Equation ~\ref{eq: vprime}, yielding the final state vector $\vec{s_f}$.  The integrator is housed in the \texttt{run\_T()} method, which processes the results and assigns the relevant outputs as attributes of the \texttt{QCT} class. 

The \texttt{hamiltonian()} method defines the Hamiltonian of the system, and outputs the total energy and momentum, useful for checking conserved quantities.

\subsection{Output}
For a single trajectory, the outputs are:
\begin{enumerate}
    \item Product count $\vec{n_f}$ = ($n_q,n_{r_1},n_{r_2},n_d$) where $n_{q,r_1,r_2,d}$ is either $0$ or $1$. For atom-diatomic molecule scattering, the two possible reactions are represented by $n_{r_1}$ and $n_{r_2}$.
    \item Final state $\vec{s_f}$ = ($v_t, w(v',v_t), j$) where $w(v',v_t)$ is the Gaussian weight given in equation \ref{eq: gauss}. For trajectories yielding $n_d = 1$, the final state outputs $\vec{s_f}$ = (0,0,0).
    \item Final positions $\vec{\rho_f} = $($\rho_1, \rho_2$)
    \item Final momenta $\vec{P_f} = $($\vec{P}_1,\vec{P}_2$)
    \item Final time $t_f$
\end{enumerate}  
Each trajectory is labeled by the input parameters $(E_c,b)$. 
% There are two choices for output format:
% \begin{enumerate}
%     \item Short: The program aggregates the data into a file in the form:
%     $E_c, b, N_q, N_r, N_d$. Here $N_{q,r,d} = \sum^{N_t}n_{q,r,d}$, with $N_t$ the total number of trajectories run at the given $(E_c,b)$
%     \item Long: The program outputs a new line for each trajectory, containing output items 1-5 described above. They are also labeled by the input parameters ($E_c, b, v, j$) and the randomly generated parameters necessary to recreate the simulation: $d,\theta,\phi,\eta,$ described in section \ref{sec: icond}. This format is useful for inspecting individual trajectories, or mapping outputs to inputs.
% \end{enumerate}
\subsection{Parallel Implementation \& Performance}
\begin{figure}
    \centering
    \includegraphics[width = .7\textwidth]{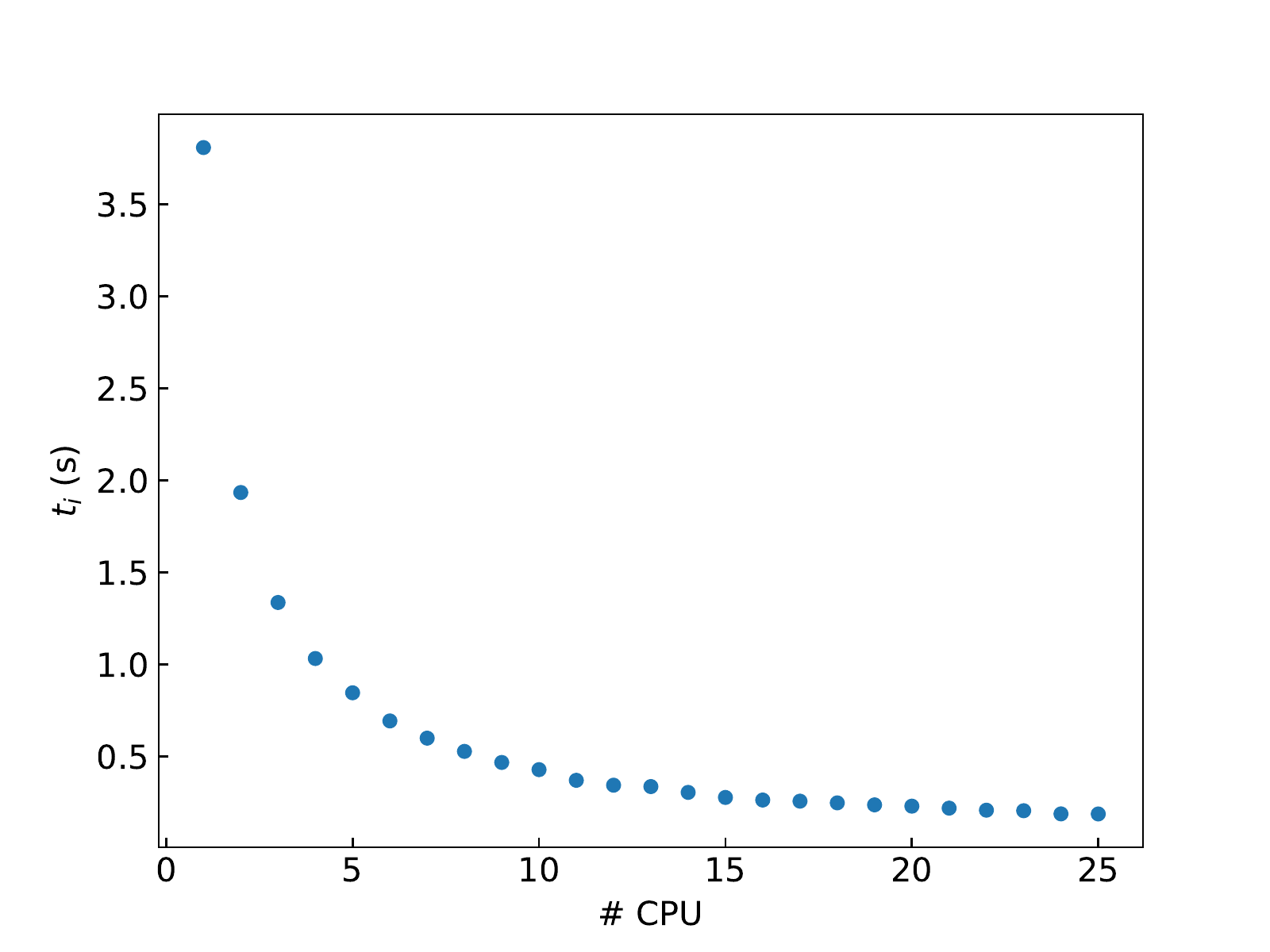}
    \caption{Time per trajectory as a function of the number of CPUs used in the parallel calculation. These trajectories were run at a collision energy of 40000 K for H$_2$ + Ca reactions, with a relative tolerance of $10^{-12}$ and absolute tolerance of $10^{-11}$.}
    \label{fig:parallel}
\end{figure}

The code is best used in a parallel implementation, which dramatically speeds up the time per trajectory as the number of CPUs is increased (Figure \ref{fig:parallel}. The \texttt{save\_short()} and \texttt{save\_long()} methods from \texttt{util.py} file uses the Python package \texttt{multiprocess} to do this. As shown in the \texttt{sim.py} file, the user can choose to save a long output, where the program outputs a new line per trajectory containing all of the output data, or a short output, which creates a new line per ($E_c$, $b$) by summing the product count vector $\vec{n_f}$. The long output is required for final state distribution analysis. 
Figure \ref{fig:parallel} shows the time per trajectory as the number of CPUs varies, averaged over 1000 trajectories. This was performed on Stony Brook's Seawulf cluster. % Not sure how to cite
It is clear that parallel processing yields a significant decrease in the total time. The majority of the calculation time is spent during the solution of the Hamilton's equations of motion. Note that for parallel processing, the multiprocess package is required. This package uses \texttt{dill}, which gives more flexibility in what can be serialized for parallel computation. 
The \texttt{solve\_ivp} API allows the user to control absolute and relative tolerances to control local error estimates. These can be controlled via the input file, and will have a large influence on the energy conservation and time per trajectory. It is highly recommended to study the effect that these tolerances have on a system before running large calculations. Each QCT object has a \texttt{delta\_e} attribute which yields the total change in energy over the trajectory. 

\subsection{Visualization}
The file \texttt{plotters.py} contains several methods for visualizing the results of a trajectory. Fig. \ref{fig:trajectories} was obtained using the \texttt{traj\_plt} function, which requires a trajectory object as input.
There is also a 3d plot generator \texttt{traj\_3d} which provides a trace of the event. Finally, there is a method to generate an animation of the trajectory, \texttt{traj\_gif}. 
The usage of these plotters is shown in the example Jupyter notebook. 

\section{Example}\label{example}
As an example, we demonstrate the calculation of CaH formation rate as a result of the reaction H$_2$ + Ca $\rightarrow$ CaH + H. The input file inputs.json. All of the simulation and potential parameters are listed. The potential ranges and parameters for H$_2$ and CaH were obtained from \cite{h2_par,h2_pec} and \cite{cah_pec,cah_par}, respectively. We include the parameters of the three potentials we might be interested to study this system in; Morse, Lennard-Jones ('lj'), and Buckingham ('buck'). Here we choose the Morse potential to describe the interaction of both H$_2$ and CaH: 
\begin{equation}
    V(r) = D_e\left(1-\exp^{(-\alpha(r-r_e))}\right)^2 - D_e
\end{equation}
% \begin{figure}[htp]
%     \centering
%     \includegraphics[scale = 0.7]{figures/ex_potential.png}
%     \caption{Morse potential describing H$_2$ and CaH interactions}
%     \label{fig:morse}
% \end{figure}

In this example, we run $10^4$ trajectories in parallel, looped over 20 impact parameters. (Listing \ref{code: parallel}) \\ 
\lstset{language=Python,
        basicstyle=\fontsize{9}{11}\ttfamily,
        keywordstyle=\color{blue}\ttfamily,
        commentstyle=\color{green}\ttfamily,
        stringstyle=\color{red}\ttfamily,
        breaklines=True
}

\begin{lstlisting}[label = {code: parallel},
                   caption = {Parallel implementation of PyQCAMS with comments. The ``utils.save\_short()" runs the quasiclassical trajectories in parallel and outputs the summed data to the specified ``out\_file".}]
import pyqcams.pymar as pymar
import numpy as np
import pyqcams.util as utils
import os

if __name__ == '__main__':
    calc = pymar.start('inputs.json') # Calculated parameters for main function

    n_traj = 10000 # Number of trajectories
    out_file = f'example/results_short.csv'
    cpus = os.cpu_count() # Number of cpus for parallel calculation
    bvals = np.arange(0,5,.25) # Range of impact parameters

    # loop over all impact parameters
    for b in bvals:
        calc['b'] = b # set new impact parameter
        # Uncomment below for long output
        # utils.save_long(n_traj, cpus, calc, f'{out_file}') 
        # Uncomment below for short output
        utils.save_short(n_traj, cpus, calc, f'{out_file}') 
    

\end{lstlisting}

\begin{figure}[ht]
    \centering
    \includegraphics[width = .8\textwidth]{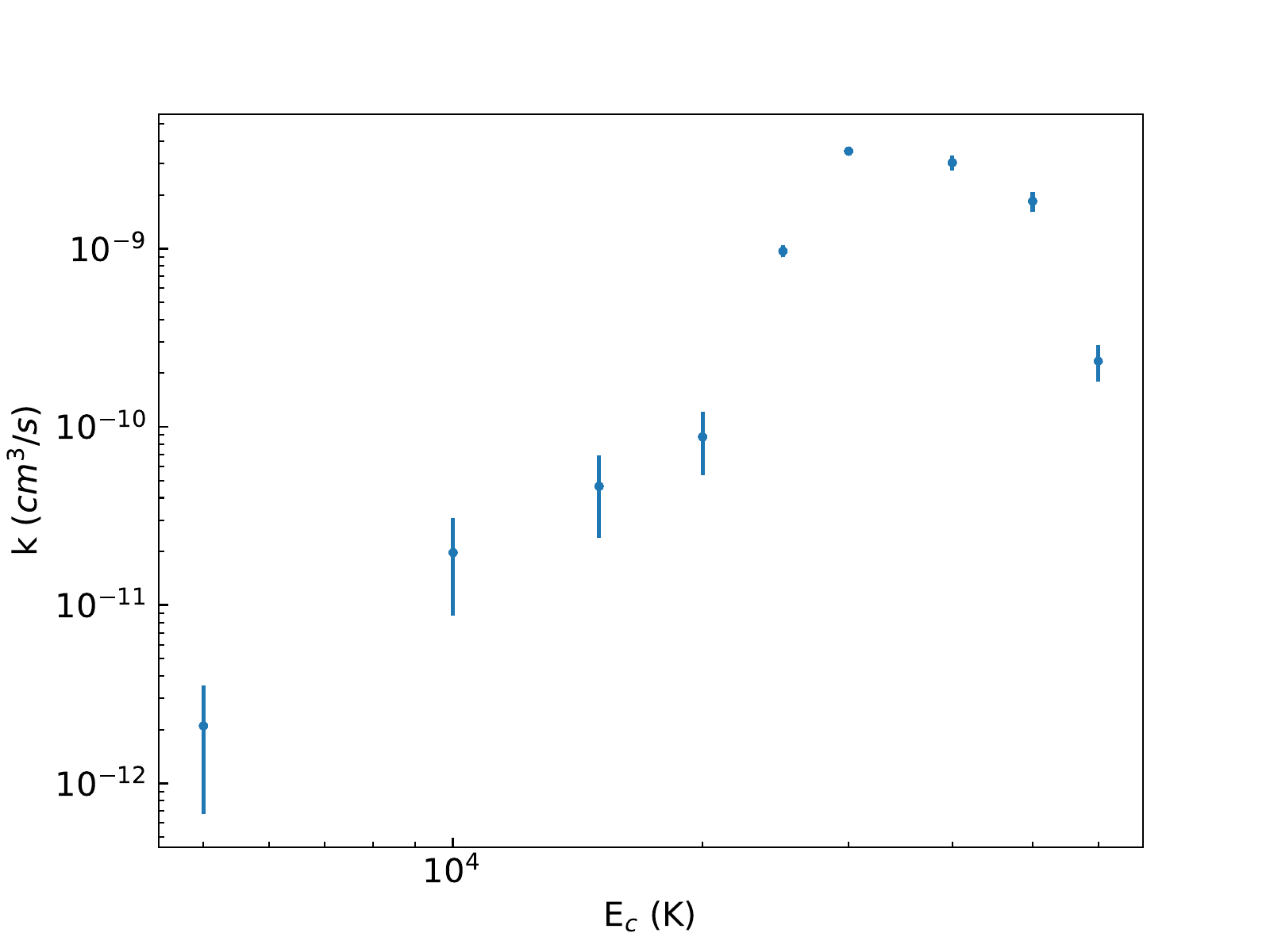}
    \caption{Rate of reaction H$_2$ + Ca $\rightarrow$ CaH + H. 9 different collision energies were considered, and $10^4$ trajectories were run at each energy. We use 20 evenly spaced impact parameters between 0 and 5~a$_0$ for each collision energy. Here, the H$_2$ molecule was initiated at $v = 1$, $j = 0$ and each pairwise interaction was defined by a Morse potential. From here, we see that CaH is most likely formed at a collision energy $E_c = 30000$ K.} 
    \label{fig:rate}
\end{figure}

The output data now has $\sim$ 10000 $\times$ 20 lines, each corresponding to one trajectory. Each of the 20 impact parameters leads to a different opacity function for a reaction, $P_r(E_c,b)$. The opacity function yields the scattering cross section at different collision energies $E_c$:
\begin{equation}\label{eq:opacity}
    \sigma_{q,r,d}(E_c) = 8\pi^2\int_{0}^{b^{q,r,d}_{max}}P_{q,r,d}(E_c,b)bdb
\end{equation}
and the rate
\begin{equation}\label{eq:rate}
    k_{q,r,d}(E_c) = \sigma_{q,r,d}(E_c)\sqrt{\frac{2E_c}{\mu}}
\end{equation}
For this example, we repeated the code in Listing \ref{code: parallel} over 10 different collision energies $E_c$. For each $E_c$, we sum over the count vector $\vec{n}$ to obtain the opacity function, cross section, and rate as described. The rates for CaH formation are shown in Figure \ref{fig:rate}, where it is noticed that higher collision energies gives rise to larger reaction rates, as expected in endothermic reactions. However, at very high collision energies such a trend changes due to the dominance of molecular dissociation processes. A more detailed study of this reaction will be published elsewhere. 

\begin{figure}[ht]
    \centering
    \includegraphics[width = \textwidth]{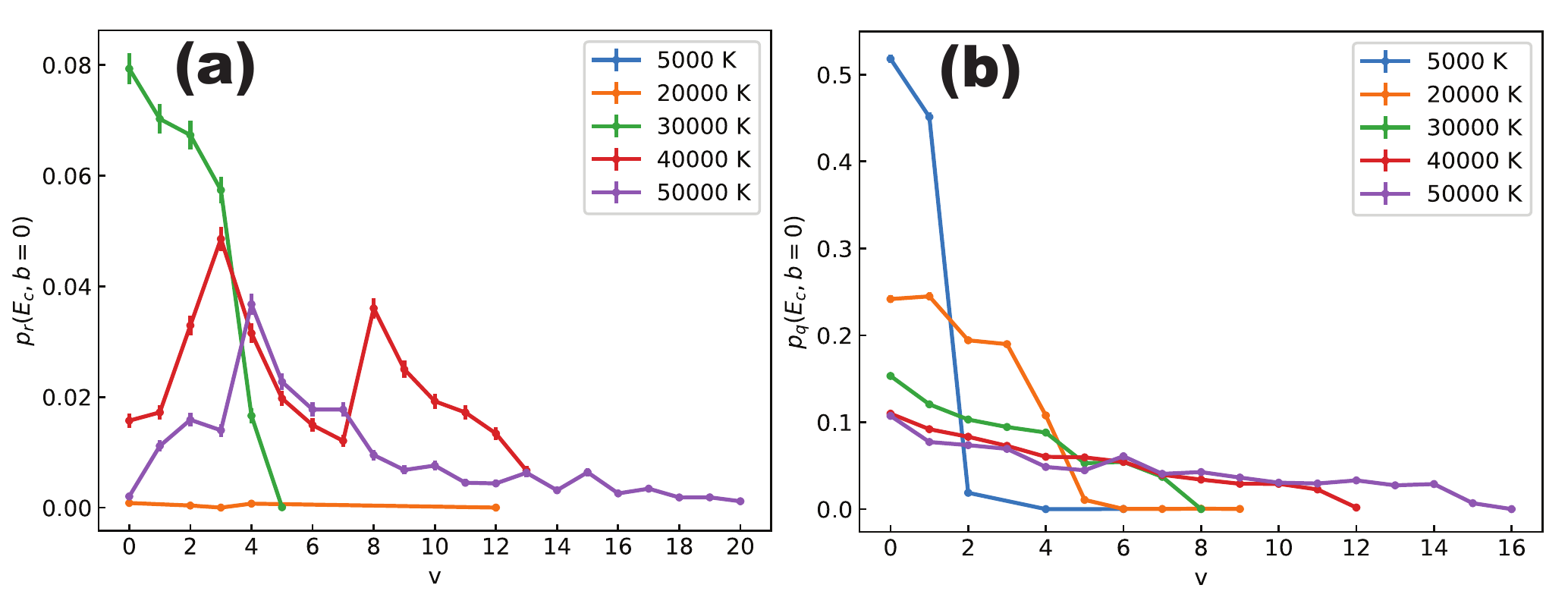}
    \caption{(Color online.) Probability distribution of final vibrational states of CaH (a) and H$_2$ (b) as a result of the reaction H$_2$ + Ca, calculated by PyQCAMS. H$_2$ was initiated at $v = 1$, $j = 0$, with each pairwise interaction defined by a Morse potential. Each color represents a different collision energy $E_c$, and the impact parameter of each trajectory is fixed to $b=0$. }
    \label{fig:distribution}
\end{figure}

We can also calculate the distribution of states of CaH and H$_2$, using the final state vector $\vec{s}$. Figure \ref{fig:distribution} shows these distributions at four different collision energies. We can see that the higher vibrational states fill up for both CaH and H$_2$ as the collision energy is increased, as is typical in endothermic reactions.  
All details of these calculations can be found in a Jupyter notebook example file. 
%\clearpage
\section{Conclusions}\label{conclusion}
We have presented a Python quasi-classical atom-molecule scattering program, PyQCAMS. The PyQCAMS program aims to provide an easy-to-use platform for calculating quasi-classical trajectories for atom-diatomic molecule systems, including the three most relevant potentials for diatomic molecules: Morse, Buckingham and the generalized Lennard-Jones.  We discussed the underlying theory behind the program, and the methods of the program as they pertain to the theory. As output, the user obtains the reaction probability per energy and impact parameter. Then, with this information is possible to calculate the cross section and energy-dependent rate constant. Its object-oriented approach allows the user to study different properties of a given trajectory. The plotter tools make it easy to visualize the results of a trajectory, making it ideal for a new researcher studying trajectories or for presenting the topic in a classroom setting. 

\section{Acknowledgments}

The authors acknowledge the generous support of the Simons Foundation. 


%% References


\nocite{*}
\bibliographystyle{elsarticle-num}
\bibliography{main}






% \end{thebibliography}

\end{document}